\documentclass[noshowpacs, amsmath, amssymb, amsthm, aps, jmp, thmsb, preprint]{revtex4}
\usepackage{color,amsbsy,amssymb,latexsym,amsfonts, amsmath,cancel}
\usepackage{mathrsfs}
\usepackage{graphicx}
\usepackage{subfigure}

\begin{document}

\title{Uncertainty Relations for Approximation and Estimation}

\author{Jaeha Lee$^1$}
\email[Email:]{jlee@post.kek.jp}
\author{Izumi Tsutsui$^{1, 2}$}
\email[Email:]{izumi.tsutsui@kek.jp}

\affiliation{$^1$Department of Physics, University of Tokyo, 7-3-1 Hongo, Bunkyo-ku, Tokyo 113-0033, Japan\\
$^2$Theory Center, Institute of Particle and Nuclear Studies,
High Energy Accelerator Research Organization (KEK), 1-1 Oho, Tsukuba, Ibaraki 305-0801, Japan
}

\date{\today}
\begin{abstract}
We present a versatile inequality of uncertainty relations which are useful
when one approximates an observable and/or estimates a physical parameter based on the measurement of
another observable.  It is shown that the optimal choice for proxy functions used for the approximation
is given by Aharonov's weak value, which also determines the classical Fisher information in parameter estimation, 
turning our inequality into the genuine Cram{\'e}r-Rao inequality.   Since the standard form of the uncertainty relation
arises as a special case of our inequality, and since the parameter estimation is available as well, our inequality
can treat both the position-momentum and the time-energy relations in one framework albeit handled differently.

\end{abstract}

\maketitle

\section{introduction}

Uncertainty relations lie undoubtedly at the heart of quantum mechanics, characterizing the indeterministic
nature of microscopic
phenomena which stem from the incompatibility of simultaneous measurement of two non-commuting observables, as typically exemplified by
position and momentum.
Soon after the celebrated exposition of Heisenberg's tradeoff relation between error and disturbance \cite{Heisenberg}, there
appeared a revised form called the Robertson-Kennard (RK) inequality \cite{Kennard, Robertson} which refers to 
the relation in standard deviation in independently performed measurements on the two observables.  
Because of its mathematical clarity and universal validity, the latter has now become a standard textbook material.

Later, these relations were elaborated from operational viewpoints by taking account of the measurement device, and this has yielded, {\it e.g.}, 
the Arthurs-Goodman inequality \cite{Arthurs-Goodman} and the Ozawa inequality \cite{Ozawa} which concerns a mixed 
relation among error, disturbance and standard deviation.   
Apart from these, the uncertainty relations on error and disturbance have also been analyzed in quantum estimation theory \cite{Watanabe}.
 
On the other hand, the uncertainty relation between time and energy has to be dealt with quite independently from these, 
due to the lack of a genuine time operator conjugate to the Hamiltonian.   For this, several ingenious frameworks have been proposed, including 
the one devised by Mandelshtam-Tamm \cite{Mandelshtam-Tamm} and that by Helstrom \cite{Helstrom}, where
the uncertainty relation is shown to be identified with a quantum version of 
the Cram{\'e}r-Rao inequality \cite{Cramer} in estimation theory.  

In this paper, we present a novel inequality of uncertainty relations for analyzing the error of approximating
an observable based on the measurement of another observable through an appropriate choice of proxy functions.  
Since the standard deviation may be regarded as a special case of our approximation error, our inequality can formally be considered as an extension of
the RK inequality.  Moreover, 
instead of approximating an observable, 
we may also choose to estimate a physical parameter pertinent to the observable, so that
the time-energy relation can be treated along with the position-momentum relation.

Interestingly, in both approximation and estimation, Aharonov's weak value \cite{Aharonov} of the concerned observable arises as
a key geometric ingredient, deciding the optimal choice for the proxy functions.  We shall also find 
in the context of parameter estimation that the weak value determines the classical Fisher information and turns
our inequality into the Cram{\'e}r-Rao inequality.

\section{Uncertainty relation for approximation}

Before presenting our uncertainty relation, 
let us recall the most familiar form of the relations, {\it i.e.}, the Robertson-Kennard (RK) inequality,
\begin{equation}
\label{ineq:Robertson-Kennard}
\|A - \langle A \rangle\| \cdot  \|B - \langle B \rangle\| \geq {1 \over 2} \left|\, \left\langle [A,B] \right\rangle\,\right|,
\end{equation}
valid for two observables $A$, $B$.  Here, $ \langle A \rangle = \langle \psi| A |\psi\rangle$ is the expectation value of $A$ under a given (normalized) state $|{\psi}\rangle$, $[A,B] = AB - BA$ is the commutator, and 
$\|X\| = \sqrt{\langle  X^2 \rangle}$ is the operator seminorm defined for a self-adjoint operator $X$.   
The RK inequality gives the lower bound for the product of the standard deviation $\sigma(X) = \sqrt{\hbox{Var}[X]} = \|X - \langle X \rangle\|$ 
for the two observables $A$ and $B$.  
Needless to say, the lower bound of the RK inequality \eqref{ineq:Robertson-Kennard} takes the state-independent value $\hbar/2$
when $A$ and $B$ are canonically conjugate to each other, $[A, B] = i\hbar$.

In place of the standard symmetric treatment of the observables in the RK inequality \eqref{ineq:Robertson-Kennard}, we now consider an asymmetric but more versatile form given by
\begin{equation}
\label{ineq:new}
\|A - f(B)\| \cdot \| g(B) \|  \geq {1 \over 2} \left|\, \left\langle [A,  g(B)] \right\rangle\,\right|,
\end{equation}
valid for arbitrary self-adjoint operator functions $f(B)$ and $g(B)$ of $B$ (the advantage of this asymmetric treatment will become clear shortly).  More explicitly, the operator $f(B)$ is defined 
from a function $f(b)$ of $b$ through the spectral decomposition,
\begin{equation}
\label{eq:opfunc}
f(B) = \int f(b) | b\rangle \langle b | \,db,
\end{equation}
where $\{| b\rangle\} $ is the basis set of the eigenstates $B| b\rangle = b| b\rangle$, and
the integral in \eqref{eq:opfunc} is understood to imply summation when the eigenvalues are discrete.   
The operator $g(B)$ can be defined analogously from a function $g(b)$. In our discussion, $f(b)$ and $g(b)$ are assumed real so that both $f(B)$ and $g(B)$ are self-adjoint, but the definition \eqref{eq:opfunc}
can be applied to any normal operators using complex functions.    The RK inequality 
\eqref{ineq:Robertson-Kennard} arises from \eqref{ineq:new} as a special case by letting $f(B) =  \langle A \rangle$ (realized by the constant function $f(b) = \langle A \rangle$) and $g(B) = B -\langle B\rangle$.

The proof of the inequality \eqref{ineq:new} goes precisely the same way as that of the RK inequality.  Namely, 
given two self-adjoint
operators $X$, $Y$, we have $\| X\|^2 \cdot \|Y \|^2 \geq |\langle XY \rangle |^2$ by the Cauchy-Schwarz (CS) inequality.   We also have
$|\langle XY \rangle |^2 =  |\langle [X, Y]/2 +  \{X, Y\}/2\rangle |^2$ where $\{X, Y\} = XY+ YX$, but 
since $\langle [X, Y]/2\rangle$ is purely imaginary whereas $\langle\{X, Y\}/2\rangle$ is real, we obtain 
$|\langle XY \rangle |^2 = |\langle [X, Y]/2\rangle |^2 +   |\langle\{X, Y\}/2\rangle |^2 \geq |\langle [X, Y]/2\rangle |^2$.
Combining these, and taking the square root of the two sides, we arrive at
$\| X\| \cdot \|Y \| \geq   {1 \over 2} \left|\, \left\langle [X, Y] \right\rangle\,\right|$.
Since $X$ and $Y$ are arbitrary, we may put $X = A - f(B)$ and $Y = g(B)$ 
to obtain our inequality \eqref{ineq:new}.  

Although our inequality is merely an asymmetric generalization
of the RK inequality, the acquired form \eqref{ineq:new} 
allows for a novel viewpoint on the uncertainty relation.  Specifically, noting that $\|A - f(B) \| $ gives a measure for the \lq distance\rq\ between
the two observables $A$ and $f(B)$, we may regard \eqref{ineq:new} as
an inequality giving the lower bound for the distance under the choice of $f(B)$ and $g(B)$.
This will be made more apparent by introducing $\bar g(B) = g(B)/\|g(B)\|$ to rewrite \eqref{ineq:new} as
\begin{equation}\label{eq:new_uncertainty_alt}
\min_{f} \|A - f(B) \| \geq  \max_{\bar g} {1 \over 2} \left|\left\langle [A, \bar g(B)] \right\rangle \right|.
\end{equation}
This indicates that the minimal distance between $A$ and the family of all self-adjoint operators $f(B)$ generated by $B$, or the minimal error in the approximation of $A$ from the measurement of $B$ in terms of real proxy functions $f(b)$, is dictated by the maximal degree of non-commutativity of $A$ with respect to the family of all self-adjoint operators $\bar g(B)$ normalized as $\| \bar g(B) \| = 1$.

Clearly, our inequality will be useful in the operational context in which 
one measures only $B$ and approximates $A$
out of the measurement result by choosing the proxy function $f(B)$ properly.  
In this context, the choice $f(B) =  \langle A \rangle$, which 
makes the distance into the standard deviation $\sigma(A)$, is far from the optimal one, having only its expectation value right.
In fact, we shall see shortly that 
the optimal choice for $f(b)$ is provided explicitly by the real part of the weak value $A_{w}(b)$ which is defined in the quantum process specified by 
the initial state $| \psi\rangle$ and the final state $| b\rangle$ (see \eqref{eq:wv}).  
Under this optimal choice together with $g(B) = B -\langle B\rangle$, our inequality \eqref{ineq:new} yields an uncertainty relation stricter than the RK inequality.  The freedom of choice for $g(b)$ may further be exploited for considering parameter estimation, that is, for estimating a parameter $t$ that specifies the state,
as exemplified later by the situation in which the state varies unitarily with the generator $A$.  In this context, the optimal choice for $g(b)$ turns out to be given by the imaginary part of the weak value $A_{w}(b)$.  

\section{Optimal Choice and the Weak Value}

In what follows, we assume for simplicity the non-degeneracy of $B$ and the condition $\langle b | \psi\rangle \ne 0$ for all $| b \rangle$, which can always be ensured if one chooses $B$ appropriately with respect to the given state $| \psi \rangle$.  The primary role of this restriction is to avoid the mathematical elaborations required to introduce the weak value
\begin{equation}
\label{eq:wv}
A_{w}(b)= {{\langle b | A | \psi\rangle}\over{\langle b | \psi\rangle}}
\end{equation}
in a completely general manner, so that we may regard it simply as a function $b \mapsto A_{w}(b)$ from the eigenvalues of the self-adjoint operator $B$ to complex numbers.  (In order to lift this restriction, we need to adopt the definition of the weak value being an equivalence class of the family of functions with ambiguity at the singular points $\langle b | \psi\rangle = 0$, for which our argument below goes through without an essential change.)  Here, note also the state-dependence of the weak value $A_{w}(b)$.

Now, in order to see the statements made above, let us first note the identity,
\begin{align}
\label{id:importantid}
A | \psi \rangle 
&= \int  | b\rangle \langle b | A | \psi\rangle \,db \nonumber \\
&= \int  A_{w}(b) | b\rangle \langle b | \psi\rangle \,db  
=  A_{w}(B)| \psi \rangle,
\end{align}
where 
\begin{equation}
A_{w}(B) = \int A_{w}(b) | b\rangle \langle b | \,db
\end{equation}
is the operator function defined analogously to \eqref{eq:opfunc} with $f(b)$ replaced formally by the weak value $A_{w}(b)$ given in \eqref{eq:wv}.  Note that, due to the complex nature of $A_{w}(b)$, the operator $A_{w}(B)$ is not necessarily self-adjoint but rather normal, and that it is dependent on the choice of the state $|\psi\rangle$.

Then, from \eqref{id:importantid} we have
\begin{equation}
\label{id:af}
\langle f(B) A \rangle = \langle f(B) A_{w}(B)\rangle
\end{equation}
for any self-adjoint $f(B)$, and thus
\begin{align}
\label{id:correl}
\begin{split}
\mathrm{Re}\, \langle f(B) A \rangle & = \langle f(B) \mathrm{Re}A_{w}(B)\rangle, \\
\mathrm{Im}\, \langle f(B) A \rangle  &= \langle f(B) \mathrm{Im}A_{w}(B)\rangle,
\end{split}
\end{align}
which specifically leads to
\begin{align}
\label{eq:rewvid}
\begin{split}
\langle A \rangle & = \langle \mathrm{Re} A_{w}(B) \rangle, \\
0  &= \langle \mathrm{Im}A_{w}(B)\rangle
\end{split}
\end{align}
for the choice of the constant function $f(b) = 1$.
Another consequence of \eqref{id:importantid} is 
\begin{equation}
\label{eq:asquare}
\|A \|^2 =  \|\mathrm{Re} A_{w}(B)\|^2 +\| \mathrm{Im} A_{w}(B)\|^2.
\end{equation}

\begin{figure}
\includegraphics[width=160mm]{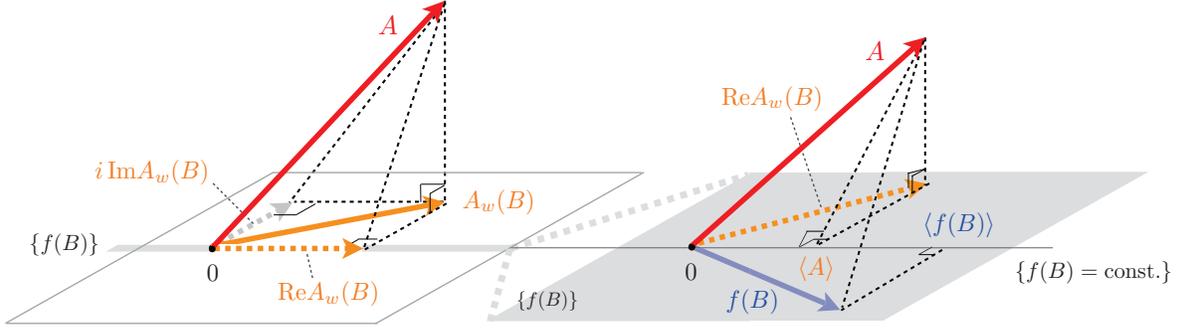}
\caption{Geometric relations among the operators involved.  The right angle symbol indicates the validity of the `Pythagorean identity'. The left illustrates how the operator $A$ is `projected' onto the subspace of normal operators generated by $B$, with the center line representing the space of self-adjoint operators $\{ f(B)\}$.   The right elaborates the projection onto the space $\{ f(B)\}$, where
now the center line represents the space of constant functions $\{ f(B) = \hbox{const.} \}$ (more precisely, functions proportional to the identity operator $I$) including $f(B) = \langle A\rangle$.
}
\label{fig:geometry}
\end{figure}

These statistical properties on average and correlation suggest that the operator $\mathrm{Re} A_{w}(B)$ may furnish the optimal proxy function for $A$
minimizing the distance $\|A - f(B)\|$.
That this is indeed the case can be confirmed at once from the `Pythagorean identity'
{
\begin{equation}
\label{eq:optimal}
\|A - f(B)\|^2
 = \| A -  \mathrm{Re} A_{w}(B)\|^2 + \| \mathrm{Re} A_{w}(B) - f(B)\|^2
\end{equation}
}%
valid for the pair of operators
$A -  \mathrm{Re} A_{w}(B)$ and $\mathrm{Re} A_{w}(B) - f(B)$, whose `orthogonality'
\begin{equation}
\mathrm{Re}\, \langle \,(A -  \mathrm{Re} A_{w}(B))\, (\mathrm{Re} A_{w}(B) - f(B))\,\rangle
    = 0,
\end{equation}
may be readily confirmed by a straightforward application of \eqref{id:correl}.

We thus see that the optimal choice for $f(B)$ is made by
\begin{equation}
\label{eq:fopt}
f_{\rm opt}(B) = \mathrm{Re} A_{w}(B)
\end{equation}
for which we have $\langle f_{\rm opt}(B) \rangle = \langle A \rangle$ as expected \cite{Hall_2001,Johansen}.
As for the optimal choice for $\bar g(B)$ attaining the maximal value for the commutator in the r.h.s.~of \eqref{eq:new_uncertainty_alt}, one readily learns from
the equality condition of the CS inequality applied to the identity
\begin{align}
{1 \over 2} \left|\left\langle [A, \bar g(B)] \right\rangle \right| = \left|\left\langle \overline{g}(B) \mathrm{Im}A_{w}(B) \right\rangle \right|
\end{align}
that $\bar g_{\rm opt}(B) = \mathrm{Im} A_{w}(B)/ \| \mathrm{Im} A_{w}(B) \|$
provides the answer.    It is then obvious that these optimal choices, $f_{\rm opt}(B)$ and $\bar g_{\rm opt}(B)$, realize the equality in \eqref{eq:new_uncertainty_alt}.  See figure~\ref{fig:geometry} for the intuitive visualisation of the geometric relations among the operators involved.

\section{RK Inequality Revisited}

Now, under the optimal choice \eqref{eq:fopt} for $f(B)$, one may put $g(B) = B - \langle B \rangle$ in \eqref{ineq:new} to obtain
\begin{equation}
\| A -  \mathrm{Re} A_{w}(B)\| \cdot \| B - \langle B \rangle \|  
\geq {1\over 2} \left|\, \left\langle [A,   B ] \right\rangle\,\right| .
\label{ineq:imaginary} 
\end{equation}
Recalling that the RK inequality arises at the non-optimal choice 
$f(B) = \langle A \rangle$, we see that, apart from the trivial case where the l.h.s.~vanishes, the inequality \eqref{ineq:imaginary}
is tighter than the RK inequality \eqref{ineq:Robertson-Kennard}.
It is also evident that \eqref{ineq:imaginary} reduces to the RK inequality if
\begin{equation}
\label{eq:eigencond} 
\mathrm{Re} A_{w}(B) | \psi \rangle =  \langle A\rangle | \psi \rangle,
\end{equation}
in which case the covariance,
\begin{align}
\label{eq:covident}
\hbox{Cov}[A, B] 
&={1\over 2}\left\langle \{A, B \} \right\rangle - \langle A \rangle\langle  B  \rangle \nonumber \\
&= \langle \left(\mathrm{Re} A_{w}(B) - \langle A \rangle\right)  \left(B - \langle B \rangle\right) \rangle,
\end{align}
vanishes identically.  

An elementary example to illustrate our point is provided by the 1-qubit system with $A = \sigma_{x}$, $B = \sigma_{z}$.   Writing 
\begin{equation}
|\psi\rangle = \begin{pmatrix}
    \phantom{e^{i\varphi}} \cos\left({\theta}/{2}\right) \\
    e^{i\varphi} \sin\left({\theta}/{2}\right)
    \end{pmatrix}, \quad 0 \leq \theta \leq \pi,\ 0 \leq \varphi \leq 2\pi,
\end{equation}
in the Bloch sphere representation, one obtains 
\begin{align}
\|\mathrm{Re}A_{w}(B) - \langle A \rangle\| = \left| \cos\theta \cos\varphi \right|,
\quad \| B - \langle B \rangle\| = \left| \sin\theta \right|.
\end{align}
One thus finds that generically (as long as
$\theta \neq 0,\, \pi/2,\, \pi$ and $\varphi \neq \pi/2,\, 3\pi/2$) 
our inequality \eqref{ineq:imaginary} gives a tighter relation than the RK inequality.

Now, notice that applying the CS inequality to \eqref{eq:covident} yields another inequality,
{
\begin{equation}
\| \mathrm{Re} A_{w}(B) -\langle A\rangle  \| \cdot \| B - \langle B \rangle \| 
\geq \left|\, {1\over 2}\left\langle \{A, B \} \right\rangle - \langle A \rangle\langle B \rangle\,\right| .
\label{ineq:real} 
\end{equation}
}%
We then see, from \eqref{id:af} and \eqref{eq:covident}, that the lower bound is equal to
$\hbox{Cov}[\mathrm{Re} A_{w}(B), B]$, and hence \eqref{ineq:real} is nothing but the classical covariance 
inequality $\sigma(X)\sigma(Y) \geq \mathrm{Cov}[X,Y]$.
This should be the case, because the operators appearing in \eqref{ineq:real} are all generated by $B$ and, accordingly, they are simultaneously measurable.

From this observation we learn that, while the inequality \eqref{ineq:imaginary} gives a purely quantum lower bound for 
the product of error in approximating $A$ and the standard deviation of $B$, 
the inequality \eqref{ineq:real} gives a classical lower bound given by the covariance
of the two observables.  
These two may be regarded as complementary to each other in view of the fact that, if we sum them 
after squaring the both, we find the Schr{\"o}dinger inequality \cite{Schroedinger},
\begin{align}
\label{ineq:Schroedinger}
&\|A - \langle A \rangle\|^2\cdot \|B - \langle B \rangle\|^2 \nonumber \\
&\quad \geq 
 \left|\, {1 \over 2}\left\langle [A,B] \right\rangle\,\right|^{2} 
+  \left|\, {1 \over 2} \left\langle \{A,B\} \right\rangle- \left\langle A \right\rangle\left\langle B \right\rangle\,\right|^{2},
\end{align}
which is a tightened version of the RK inequality.   

In passing, we observe from the equality condition of the CS inequality and the identity
\begin{equation}
\| A -  \mathrm{Re} A_{w}(B)\|= \| \mathrm{Im} A_{w}(B)\|,
\label{eq:distsq} 
\end{equation}
which follows from \eqref{id:importantid} that
the equality in \eqref{ineq:imaginary} holds if
\begin{equation}
\mathrm{Im} A_{w}(B)|\psi\rangle = \lambda (B - \langle B \rangle)|\psi\rangle,
\label{eq:imwvid}
\end{equation}
for some real $\lambda$, whereas the equality in \eqref{ineq:real} holds if
\begin{equation}
\left(\mathrm{Re} A_{w}(B) - \langle A \rangle\right)|\psi\rangle = \mu (B - \langle B \rangle)|\psi\rangle,
\label{eq:rewvid}
\end{equation}
for some real $\mu$.  Combining \eqref{eq:imwvid} and \eqref{eq:rewvid}, and using \eqref{id:importantid} again, we
obtain
\begin{equation}
\label{eq:condmu}
\left(A - \langle A \rangle\right) |\psi\rangle = \beta (B - \langle B \rangle)|\psi\rangle
\end{equation}
with $\beta = \mu + i\lambda$.  Obviously, when position and momentum are considered for the observables $A = Q$, $B = P$,
the condition \eqref{eq:condmu} reduces to the standard minimal uncertainty condition, in which $\beta$ is related to the parameter characterizing the squeezed coherent states.

\section{Parameter Estimation and Time-Energy Uncertainty Relation}

Returning to the original form \eqref{ineq:new}, we consider a family of states $|\psi(t)\rangle = e^{-itA/\hbar}|\psi\rangle$ generated by $A$ with a real parameter $t$ for a fixed $|\psi\rangle$.  Suppose that 
our aim is to find the best function $g(B)$ to estimate the parameter $t$ around a certain time $t = t_0$ by looking at the expectation value 
$\langle g(B) \rangle = \langle \psi(t)| g(B) |\psi(t)\rangle$.
In more technical terms, we wish to find
the locally unbiased estimator $g(B)$ fulfilling
\begin{align}
\label{cond:locunbiased}
\langle g(B) \rangle\vert_{t=t_0} = t_0, 
\quad
\frac{d}{dt}\langle g(B) \rangle\vert_{t=t_0}  = 1,
\end{align}
such that the variance $\hbox{Var}[g(B)]$ becomes minimal.

At this point, it is important to recognize from \eqref{eq:distsq} that the optimal value $\| \mathrm{Im} A_{w}(B)\|^2$ of
the minimal (squared) error in the approximation of $A$
that arises under $f_{\rm opt}(B)$ is nothing but the Fisher Information associated with the probability density $p(b, t) = \vert \langle b | \psi(t) \rangle \vert^2$.
Indeed, when the distribution $p(b, t)$ is regarded as the likelihood function for $t$, the corresponding Fisher Information reads
\begin{align}
I(t) 
= \int \left[\frac{d}{dt} \ln p(b, t)\right]^2 p(b, t) \,db 
= \left(\frac{2}{\hbar}\right)^2 \| \mathrm{Im} A_{w}(B)\|^2.
\end{align}
This prompts us to put $g(B) \to g(B) - \langle g(B) \rangle$ in \eqref{ineq:new} at
the optimality \eqref{eq:fopt} and see that our inequality turns into the Cram{\'e}r-Rao inequality \cite{Cramer},
\begin{equation}
\hbox{Var}[g(B)] 
\geq \frac{\left(\frac{d}{dt}\langle g(B) \rangle\right)^2}{I(t)},
\end{equation}
on account of the identity $\frac{d}{dt}\langle g(B) \rangle = \frac{i}{\hbar} \left\langle [A, g(B)] \right\rangle$.
The connection between the uncertainty relation and a quantum counterpart of the Cram{\'e}r-Rao inequality
has been pointed out earlier in estimation theory \cite{Helstrom}, but here we notice that the precise connection between our uncertainty relation and the classical Cram{\'e}r-Rao inequality holds when the optimal choice is made as also mentioned in \cite{Hofmann}.
The optimal choice of the efficient estimator $g(B)$ fulfilling \eqref{cond:locunbiased} and attains 
the lower bound is now readily given by
\begin{equation}
\label{eq:gopt}
g_{\rm opt}(B) = \frac{2}{\hbar I(t_0)} \mathrm{Im} A_{w}(B) + t_0,
\end{equation}
which is well-defined as long as the Fisher Information is nonvanishing $I(t_0) \ne 0$.

Let us specialize to the case where the unitary family of states is given by the time evolution 
generated by the Hamiltonian $H$.  The parameter $t$, which is to be estimated by $g(B)$, is then the time parameter of the evolution.  
Namely, we are estimating the \lq energy\rq\ of the system through $f(B)$ and the \lq time\rq\ of the system through $g(B)$, both based on the measurement of $B$.   
Plugging $A =H$ and letting $g(B) \to g(B) - t_0$ in our inequality \eqref{ineq:new}, and assuming that $g(B)$ is a locally unbiased estimator, we obtain
the time-energy uncertainty relation
\begin{equation}
\label{ineq:newterel}
\|H - f(B)\| \cdot \|t_0 - g(B) \|  \geq \frac{\hbar}{2}.
\end{equation}
The equality in \eqref{ineq:newterel} holds for the optimal choice of $f(B)$ and $g(B)$ given respectively in \eqref{eq:fopt} and \eqref{eq:gopt} for $A = H$.

Before closing we note that since the latter condition in \eqref{cond:locunbiased} is equivalent to
$\langle [g(B), H] \rangle = i \hbar$, 
a locally unbiased estimator $g(B)$ is required to be 
canonically conjugate to $H$ in the expectation value at least around $t = t_0$.   
The existence of such $g(B)$ is ensured from \eqref{eq:gopt} with $A = H$, which implies that an admissible form of such estimators is provided by 
$g(B) = g(B)_{\rm opt} + X$
for a time independent self-adjoint operator $X$ with $\langle X \rangle = 0$.  Of course, in the actual estimation we know neither
$H$ nor $t_0$, but at least we know that there are a host of estimators that meet our requirements.   

\section{Summary and Remarks}

To summarize, we have presented a novel inequality of uncertainty relations for approximation and/or estimation errors.
The minimal uncertainty is determined by the weak value, and in the context of estimation our inequality reduces to the Cram{\'e}r-Rao inequality.
Since out inequality contains the RK inequality as a special case, it can treat both 
the position-momentum and the time-energy uncertainty relations in one formula, even though they have to be handled differently.  
The appearance of the weak value in the relations is not accidental.  In fact, behind it lies 
the more fundamental notion of quasi-probability \cite{Lee, Mori}, whose significance in quantum mechanics should be worth exploring further.

\hspace{3mm}

This work was supported in part by JSPS KAKENHI No.~25400423,  No.~26011506, and by the Center for the Promotion of Integrated Sciences (CPIS) of SOKENDAI.



\end{document}